# Structure and Dynamics of the Inner Corona Measured from the DEB Initiative 2024 Eclipse Image Sequence

Matthew J Penn[1,2,3], Robert Baer[2], Chris Mandrell[2], Corinne Brevik[2], Castor Fu[4], Mike Conley[5], Richard Danley[2], Harvey Henson[2], David Iadevaia[6], Jonathan Mangin[7], Chris Midden[8], Brodye Miller[2], Claude Plymate[9,10], Teresa Plymate[9,10], Zack Stockbridge[11], Kevin Rasso[12], Autumn Awalt[30], Avery Awalt[13], Gracie Awalt[14], Josh Awalt[30], Lauree Rasso[30], Mike Chartrand[30], Noah Lambert[30], Zoee Rasso[15], Kevin Cobble[16], Amelia Menezes[16], Andrew Yu[16], Felix Mei[16], Hanson Du[16], Ramesh Pattar[16], Vishrut Kumaran[16], Robert Auburger[17], Brian J. Drake[30], Seth Antozzi[18], Chloe Rectanus[18], Grayson Garner[18], Michael Weiss[18], Bill Kloepping[19], Deborah Grubis[30], Maryanne Angliongto[20,21], Ashvik Chilakala[21,22], Carl Buz McCullough[21], Mark Bremer[21], Christina Adair[23], Amber Huffman[23], Charon Adair[23], Gabbie Graham[23], Nick Tillerson[23], Fred Isberner[24], Candy Isberner[24], Harry Treece[24], Lisa Sikorski[24], Sage Julian-Fralish[25], Heidi Schran[26], Addison Greene[26], Bibodh Baral[26], Crow Ely[26], Nolan Hodgson[26], Olive Blackmar Rice[26], Olivia Andrews[26], Olivia Freeman[26], Quinn Donnelly[26], Serenity Prince Kirk[26], Carmen Tran[27], Brianna Blanchard[27], Kyan West[28], Landon Bevier[27], Murali Saravanan[27], Jeremy Wright[29]

[1]Corresponding author matthew.j.penn@alumni.caltech.edu

[2]Department of Physics, Southern Illinois University Carbondale, Carbondale IL 62901, USA

[3]Tucson Amateur Astronomy Association, PO Box 41254, Tucson AZ 85717, USA

[4]Peninsula Astronomical Society, Mountain View CA 94040, USA

[5]1624 Pelican Ct NW, Salem, OR 97304, USA

[6]Mountain View Observatory (MVO) Tucson AZ

[7]Meade County High School, 938 Old State Road, Brandenburg, KY 40108, USA

[8]Unity Point Elementary School, 4033 S Illinois Av, Carbondale IL 62903, USA

[9]Bear Valley Springs Astronomy Club, 24440 Serra Place, Tehachapi, CA 93561

[10]Western Region Astronomical League, 9201 Ward Parkway Suite #100, Kansas City, MO 64114

[11]Totality Town, LLC, Andrews, NC 28901

[12]Mid-Georgia Astronomical Society, 142 Holly Pointe, Warner Robins, GA 31088

[13]Texas A&M University, 400 Bizzell St, College Station, TX 77843

[14]Midlothian Heritage High, 4000 FM1387, Midlothian, TX 76065

[15]Columbus State University, 4225 University Ave, Columbus, GA 31907

[16]Junior Texas Astronomical Society, Dallas, P.O. Box 830742, Richardson, TX 75083

[17]Towson University, 8000 York Rd, Towson, MD 21252

[18]Colorado State University Ft Collins, 200 W Lake St, Fort Collins, CO 80523

[19]Texas Astronomical Society, P.O. Box 830742, Richardson, TX 75083

[20]Jefferson College, 1000 Viking Dr, Hillsboro, MO 63050

[21]St. Louis Astronomical Society, St. Louis, MO

[22]Ladue Horton Watkins HS, 1201 S Warson Rd, St. Louis, MO 63124

[23]Southeast Missouri State University, 1 University Plaza, Cape Girardeau, MO 63701

[24]Astronomical Association of Southern Illinois, Carbondale, IL

[25]CIVA Charter High School, 4635 Northpark Dr, Colorado Springs, CO 80918

[26]Talawanda High School, 5301 University Park Blvd, Oxford, OH 45056

[27]University of Washington Seattle, 1410 NE Campus Parkway, Seattle, WA 98195

[28]North Seattle College, 9600 College Way N, Seattle, WA 98103

[29]Southern Maine Astronomers, 179 Neptune Dr, Brunswick, ME 04011

[30]Unaffiliated

## Abstract

The Dynamic Eclipse Broadcast (DEB) Initiative citizen science program observed coronal visible continuum brightness during the 2024 April 8 total eclipse from locations across North America. We present results from 11 DEB sites spanning 2700 km of distance and showing 49 minutes of evolution. The coronal brightness radial profiles from these telescopes are tightly correlated from 1.2 to 4.0 solar radii and comparable to published photometric coronal intensities. The coronal flattening parameter is measured from 1.4 to 2.8 solar radii. A Ludendorff index of 0.0761 ± 0.0007 is computed but the extrapolation techniques used by some to calculate this index are shown to disagree with this direct measurement, and alternate structure parameters are suggested. Measured radial velocities are compared with an MHD model of the corona during the eclipse from Y. Li et al. (2026). A polar downflow is measured with an average radial velocity of -37 ± 3 km $s^{-1}$ and a deceleration of 14 ± 3 m $s^{-2}$ at a speed and position which agrees with the model. The predicted mixture of outflows and downflows at low heights is seen, as well as outflows in two western regions of the corona. The fastest observed outflow has a radial speed of 105 km $s^{-1}$ and is likely associated with a transient event not predicted by the model. Future DEB Initiative eclipse experiments can more tightly constrain models of the inner corona by using both coronal intensity and radial velocity measurements.

## 1. Introduction

For more than a century, networks of observers have captured images of the inner solar corona during a total solar eclipse. An early network was established by Morton and his teams for the eclipse of 1869 August 7 (T. W. Webb 1869). Including amateur astronomers in such a network was implemented by B. H. Foing & the ESA Eclipse 99 Team (1999), and in this century H. S. Hudson et al. (2011) proposed an expanded vision of such a network. Eclipse networks with citizen scientists have produced high quality data from the 2017 total solar eclipse resulting in science results including observations of jets in polar plumes ejected at 450 km $s^{-1}$ (Y. Hanaoka et al. 2018) and measurements of 15 m $s^{-2}$ radial acceleration of CME plasma in the inner corona (M. J. Penn et al. 2020).

Building on our work from the 2017 eclipse, the Dynamic Eclipse Broadcast (DEB) Initiative established a network of 82 sites across North America using equipment and software from the amateur astronomy market to collect images of the 2024 April 8 solar eclipse. In summary our site equipment consisted of an Askar FMA 180 pro 40mm aperture 180mm focal length sextuplet Petzval-like refractor, a monochromatic Player One Neptune M camera using a Sony IMX178 CMOS array and a 14-bit ADC, an iOptron SkyHunter equatorial mount, a laptop, and a neutral density filter for non-totality observations. The Neptune M camera dictates the instrument's broad wavelength sensitivity, exhibiting an 80% peak quantum efficiency at 520nm with half-peak thresholds at 390 and 780nm. For totality observations we used SharpCap Pro to rapidly cycle between several camera exposure times to produce our high-dynamic range coronal images (C. Mandrell et al. 2025).

The weather conditions across the path of totality were not good. High cirrus clouds interfered with DEB sites in Mexico, and lower clouds compromised our images from sites in Canada. We obtained imaging data from 11 sites with clear skies that could be readily coaligned. Table 1 lists the locations and observation times of these sites. From the first clear site to the last in our network we observed 49 minutes of coronal evolution, although the time sequence does have gaps during which no coronal images were obtained. In this paper we present an analysis of coronal structure and dynamics revealed by these observations.

Table 1

DEB Observing Sites Across 2700km Baseline

| **Number** | **Site Name** | **UT** | **Site Lat** | **Site Lon** | **City State** |
|---|---|---|---|---|---|
| **1** | 39_Rasso | 18:42:00 | 32.49 | -96.90 | Midlothian TX |
| **2** | 64_Cobble | 18:43:43 | 33.16 | -96.49 | Princeton TX |
| **3** | 83_Auburger | 18:43:48 | 33.06 | -96.32 | Josephine TX |
| **4** | 20_Antozzi | 18:54:01 | 35.60 | -92.42 | Clinton AR |
| **5** | 32_Kloepping | 18:55:17 | 36.02 | -91.81 | Mt Pleasant AR |
| **6** | 19_Angliongto | 18:58:50 | 37.26 | -90.49 | Patterson MO |
| **7** | 12_Adair | 19:00:26 | 37.30 | -89.52 | Cape Girardeau MO |
| **8** | 26_Isberner | 19:01:16 | 37.60 | -89.19 | Makanda IL |
| **9** | 63_Schran | 19:09:52 | 39.49 | -84.73 | Oxford OH |
| **10** | 72_Tran | 19:10:24 | 39.66 | -84.40 | Farmersville OH |
| **11** | 33_Wright | 19:31:15 | 45.62 | -70.26 | Jackman ME |

## 2. Eclipse HDR Image Processing

The 180mm focal length telescope combined with the camera pixel size of 2.4 microns gives an angular resolution of 2.75 arcseconds per pixel, and the Dawes resolution limit for the 40mm aperture at 500nm is 3.1 arcseconds. This angular dispersion provides a nominal solar radius of 349 pixels. Eight of our 11 clear sites have images of the uneclipsed solar disk, and the average solar radius measured from these images is 345.7 +/- 0.9 pixels. This suggests there is a 1% mean error in the actual telescope focal length compared with the manufacturer specification. We use 345.7 pixels for our value of the solar radius in all of our analyses. The 3096 x 2080 pixel size of the IMX178 array gives a field of view of 2.37 x 1.59 degrees, or roughly 8.9 x 6.0 solar radii. Assuming an image of the corona is perfectly centered on the camera, we expected the white light solar coronal brightness to vary from roughly 10-6 $B_{\odot}$ at the solar limb to about 10-9 $B_{\odot}$ at the corner of our field of view. With a 14-bit camera it would be possible to capture this intensity range with one exposure, but the signal to noise ratio at the edges of our field would be very low. Instead, we took sequences of varied exposures and produced high-dynamic range (HDR) images. A sequence of HDR subframes was taken at different exposures to produce an HDR image every 5 seconds, and the collection of roughly 40 HDR images from each site was aligned and coadded to produce a summed HDR image with a higher signal to noise ratio.

To align the HDR subframes for a given site, spatial offsets were calculated using the phase correlation method described by M. Druckmuller (2009). Trinarizing the HDR subframes before implementing the phase correlation effectively removed any translation contamination from background stars and decreased the impact of noise from large radial distances in the subframes. Because we tracked the Sun with a polar-aligned mount, we had no rotational offsets among the HDR subframes from a given site. Similarly, there was no need to compute different image scale values for HDR subframes since only one optical assembly was used for all the images from a given site. While azimuthal filtering sizes varied slightly by site, they remained close to the values used by Druckmuller. Image trinarization eliminated the need for smoothing in Fourier space, but required higher additive bias values there. The neighborhood size for Fourier peak searching was identical to Druckmuller's. (In detail our analysis parameters were $\boldsymbol{\rho} \approx 8$, $\boldsymbol{\sigma} = 0$, p = q = 0.1 and $\boldsymbol{\varepsilon} = 3$.)

With all the offsets known for each HDR subframe, HDR images were produced by correlating the coronal intensities from the different subframe exposures with a process similar to M. Druckmuller et al. (2006). Since our subframe exposures differed by factors of 10 we used a larger part of the camera dynamic range to avoid spatial gaps in the resulting coronal images. Finally, all the HDR images from each site were co-added to produce one summed HDR image. Figure 1 shows a spatially filtered version of the summed HDR image from site 1.

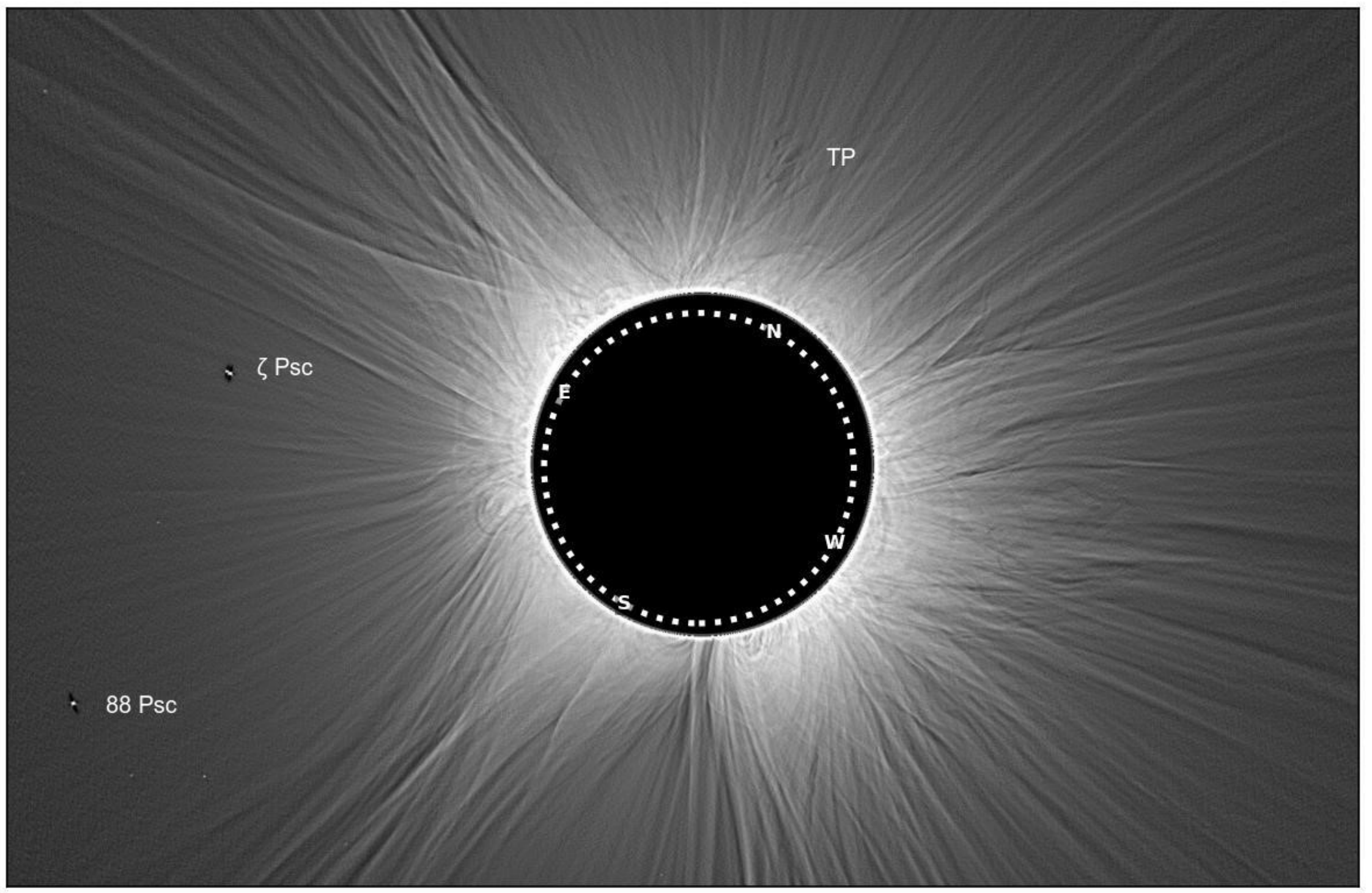

Figure 1. DEB summed HDR image from Site 1. HDR shown after filtering to reduce the radial intensity gradient and to emphasize fine structure. The solar disk position is marked with the cardinal position angle locations. Loops dominate the lowest parts of the corona and give way to more radial structures higher up. A dark downflow (TP) can be seen at about 2 solar radii just east of the north pole, a reduced coronal brightness is visible in the SE region than the western hemisphere, and the southern polar hole is very narrow. Some stars in the field-of-view are labelled, and the brightest stars show the effects of the spatial filtering on point sources.

An Affine transform matrix with zero shear was computed to map the summed HDR images from all sites to site 1, determining the required image translations, rotations, and scale factors. Stable coronal loop structures near the lunar limb provide a sanity check on the translation values returned by the transform. The Affine transform rotation values were checked by measuring the angle between the centroid of double star ς Psc A/B and location of the star 88 Psc in the summed HDR images from all sites. In all cases the rotation angles implied by the stellar positions were within 0.75 degrees of the Affine rotation angles, and the standard deviation of the differences was 0.2 degrees. Finally, the Affine image scale factors were checked in two ways. First, the solar disk radius from images taken outside of the eclipse was measured at 8 of our sites and the variations in the measured solar radius can be compared with the scale factors. The agreement is good, with a standard deviation of the differences being only 0.7%. As a second check, the observed lunar disk size was used to check the scale factors. Unlike the solar disk, the apparent size of the dark lunar disk when viewed from different sites is expected to change in size by more than 1 pixel. When the Affine image scale factors were used to calibrate the DEB lunar radius measurements, the lunar radius measurements fit the expected change within 1.1 standard deviations.

## 3. Static Coronal Structure

For each site a single parameter, the average pixel count rate from 1.7 to 2.9 solar radii, was computed and each site was normalized to the same parameter computed for Site 1. These normalization factors account for the average atmospheric and instrumental variations at each site. After making this correction, all the sites show a nearly identical decrease of total coronal intensity as a function of radial distance above the solar limb. These azimuthally averaged intensity profiles from the DEB sites are shown as the color traces in Figure 2. Above heights of about 4.0 solar radii, some alignment artifacts can be seen in this figure which appear as decreases in the averaged intensity with ($\delta I / I$) contrasts of about 0.03. One of the DEB sites deviates from the others at heights above 3.0 solar radii, while the other sites show increasing intensity differences above about 4.0 solar radii.

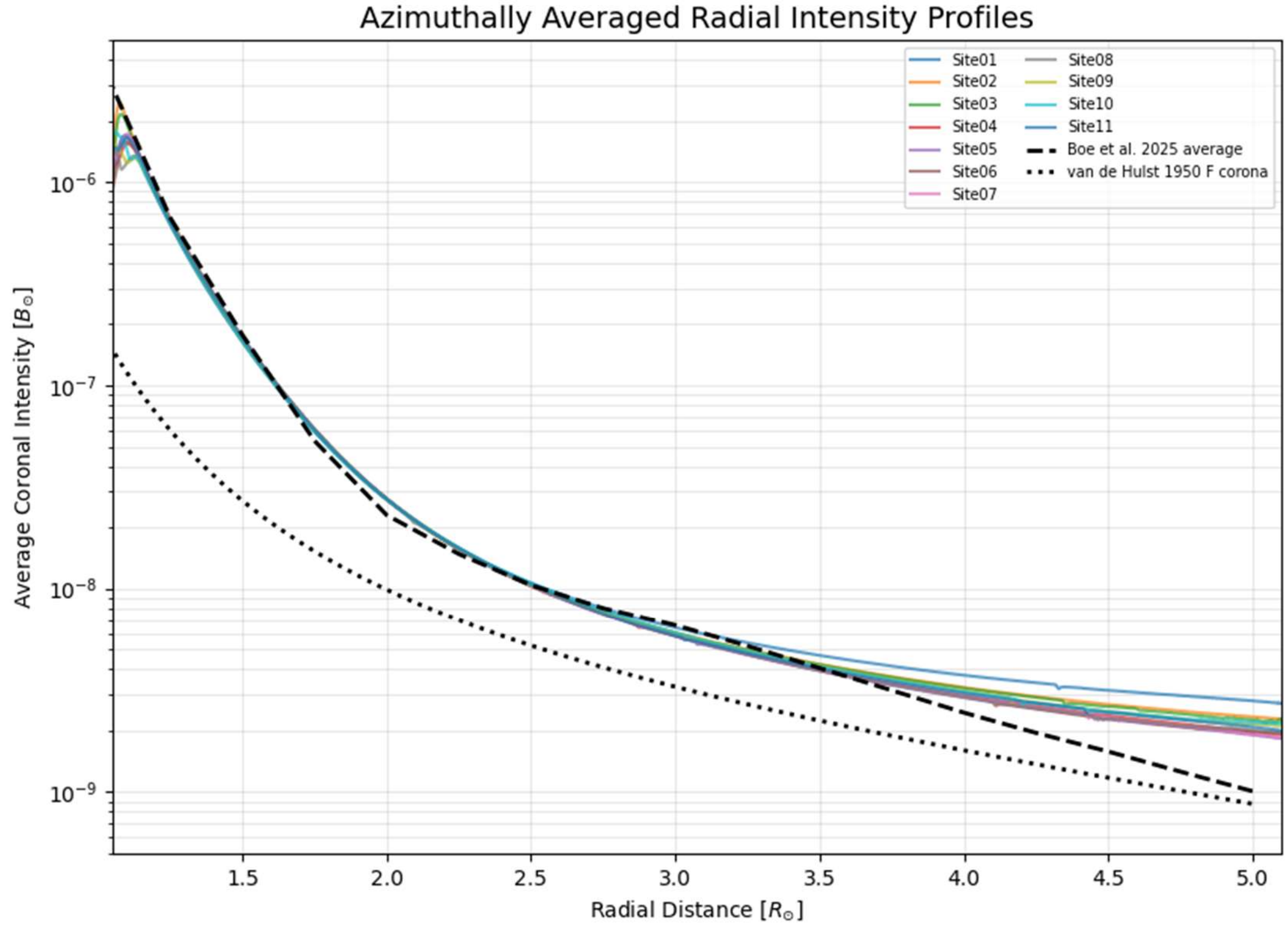

Figure 2. Radial Intensity gradients. The radial intensity gradients as measured by the summed HDR images from 11 DEB sites are shown in this plot. Intensities diverge below 1.2 solar radii because of the changing projected location of the lunar limb combined with our azimuthal averaging, and different atmospheric scattering likely causes the divergence above 4 solar radii. Small intensity drops above 4 solar radii at the 3% level are HDR subframe alignment artifacts. The radial intensity gradient from B. Boe et al. (2025) is used to normalize the DEB data to solar units and shows excellent agreement up to about 3.7 solar radii, and the H. C. van de Hulst (1950) F corona brightness is also shown. In this work we do not attempt to photometrically correct the DEB intensities above 3.7 solar radii which may suffer from atmospheric or instrumental stray light.

The DEB instrument intensities were approximately calibrated to photometric solar intensity units using the continuum profile published in B. Boe et al. (2025); their sixth figure shows continuum coronal intensity radial profiles at wavelengths near 530, 637 and 789nm along with a broadband profile (400-850nm) from the LASCO C2. The dashed line on Figure 2 shows the average intensity of all of their channels and the LASCO data at a given height in the corona. The scaled DEB profiles show good correspondence with the profile from Boe et al. up to a height of about 3.7 solar radii. While both DEB and Boe et al. coronal intensities approach the predicted F coronal value from H. C. van de Hulst (1950) at heights up to 3.7 solar radii, above this height the DEB intensities begin to diverge from the van de Hulst prediction. There is no physical reason to believe that the K corona is getting brighter at these heights, so we suggest stray light as the source for these intensity profiles. The strong radial intensity correlation across all 11 sites below 3.7 solar radii indicates negligible atmospheric or instrumental stray light, as both vary by site and equipment. Accordingly, we omit stray light corrections and restrict our structural analysis to heights below 3.7 solar radii. Intensity discrepancies below 1.2 solar radii are likely due to parallax causing differing lunar disk positions among the sites.

The solar corona shows a complicated structure. There are many dark and bright low loops visible above the lunar limb. The corona changes to become dominated by radial streamers at all position angles at the edges of the DEB field of view. In addition to polar plumes, the northern polar hole has several dark features, and the southern polar hole seems very narrow in azimuthal extent. The corona above the eastern limb shows fainter structures than the western limb. These details are clearly visible in the spatially filtered image shown in Figure 1 but are also apparent in the total intensity images from each site.

For nearly a century the structure of the solar corona during an eclipse has been characterized by the Ludendorff flattening parameter (H. Ludendorff 1928). The parameter involves measuring the diameters of total intensity isophotes along three azimuthal angles near the solar pole and three near the equator. The average isophote diameter from near the equator at position angles of 67.5, 90, and 112.5 degrees is divided by the average isophote diameter near the pole at position angles of -22.5, 0, and 22.5 degrees. From this ratio 1.0 is subtracted to obtain the flattening parameter, and the value of the parameter at 2.0 solar radii is known as the Ludendorff index and is used to characterize the corona.

The top panel of Figure 3 shows the flattening parameter as measured from 11 DEB sites, and the parameter shows a very high agreement among all telescopes. Our measured flattening does not increase linearly from 1.5 to 2.5 solar radii as described by the "rule of thumb" (M. Pishkalo 2011). Some authors (V. Rusin 2017) estimate the index value at 2.0 with a linear extrapolation using measurements at all lower heights. R. Priyatikanto (2016) first fits a parabola to measurements at many heights, determines the radial value ($R_{max}$) with the maximal flattening parameter, and then uses only measurements at $r < R_{max}$ to make a linear extrapolation to determine the index at 2.0 solar radii.

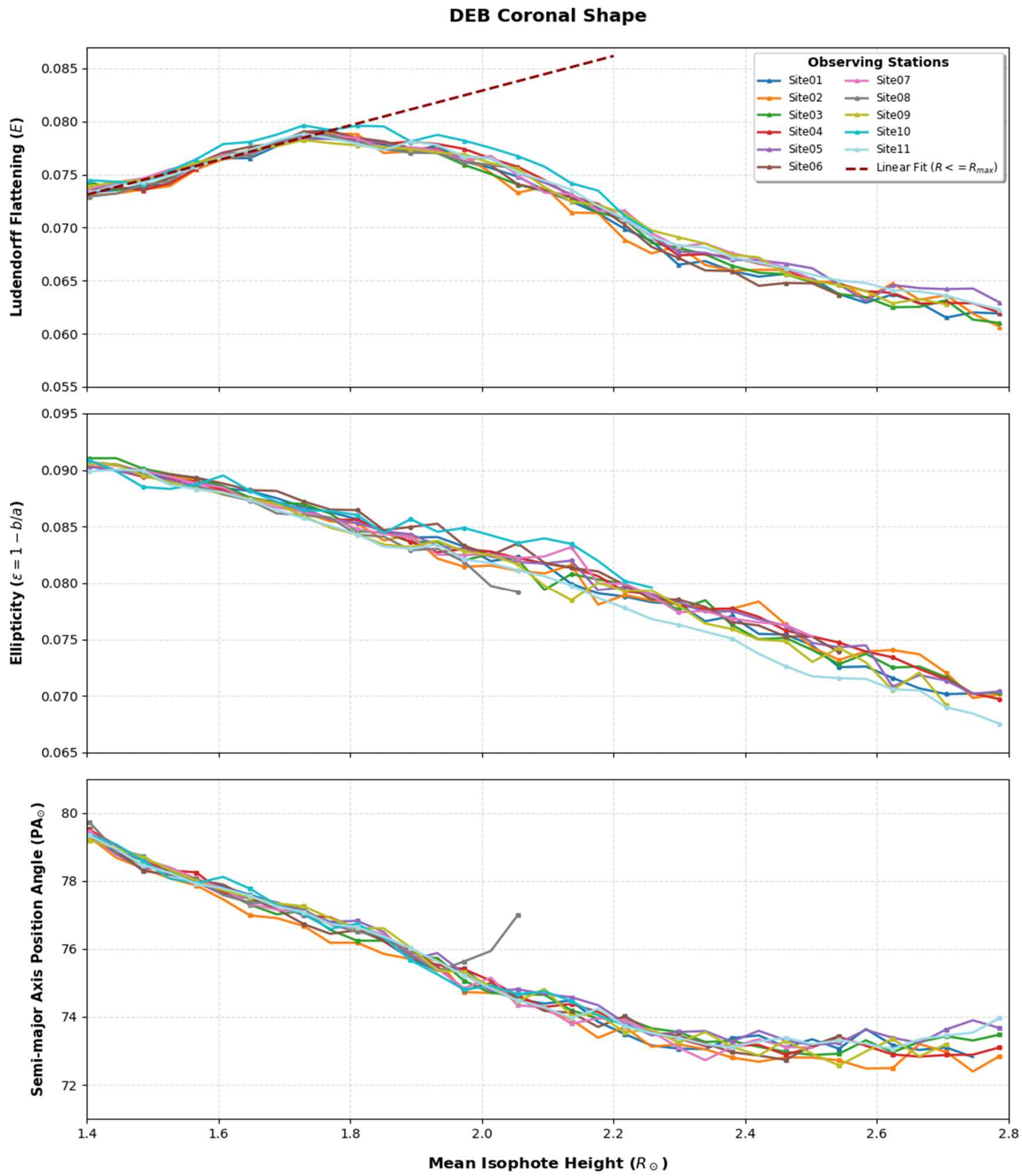

Figure 3. Ludendorff coronal flattening parameter for DEB sites. The DEB summed HDR images from all sites are used to compute the Ludendorff coronal flattening parameter as shown in the top panel. There is a high level of agreement among all the sites and the Ludendorff index at 2.0 solar radii is computed to be 0.0761 ± 0.0007. The line segment clearly results in the wrong index value and represents the extrapolation method used by some authors as discussed in the text. We also fit ellipses to the intensity isophotes and show the ellipticity and solar position angle of the semi-major axis in the second and third panels. These two parameters behave more smoothly at low coronal heights, and a linear extrapolation from lower heights would provide accurate measures at 2.0 solar radii. We suggest these parameters would better characterize the coronal structure.

Our directly measured average Ludendorff index from 11 sites at 2.0 solar radii is 0.0761 ± 0.0007. Following the parameter extrapolation technique of Priyatikanto using our data gives an index of 0.0829 ± 0.0004. Linearly fitting all of our data to compute a value does not seem warranted but would give a much lower index than our direct measurement. We conclude that the techniques used by many authors do not accurately extrapolate values of the Ludendorff index at 2.0 solar radii, and we suggest that the high degree of scatter seen in some work relating this index to solar cycle phase (i.e. the third figures in M. Pishkalo (2011) and T. Dani et al. (2016)) may be the result of the inaccuracies in these extrapolation techniques.

Can we determine a better way to characterize the coronal structure than this historic technique? In the second and third panels of Figure 3 we show plots of the ellipticity and position angle of the semi-major axis from fits to coronal brightness isophotes from all 11 sites. Isophotes which cross an image boundary are not fit, and so different sites sample different coronal heights. Both of these parameters show excellent correlation among all 11 sites, and they vary smoothly from 1.4 to 2.8 solar radii. We propose that using these two fundamental parameters from elliptical fitting of the brightness isophotes provides a more accurate way to characterize the structure of the solar corona as seen in eclipse data.

## 4. Dynamics in the Corona

The sequence of DEB coronal images shows a variety of dynamical changes during the 49 minutes from the first image to the last. The field-of-view changes and rotates for each telescope site because of the correction needed for different telescope pointing and camera rotation angles. The position of the moon relative to the solar corona also changes because of the parallax caused by having the observing sites located north and south of the eclipse centerline. The background stars move relative to the solar corona because of Earth’s orbital motion.

Regarding changes in the solar corona: low loops seen near the limb are mostly constant, showing little change during our image sequence. Just east of the north pole a "tadpole-like" downflow event is seen. Near the east limb material moves outward and spirals slightly. This is the region with the strongest observed radial outflows, and while this event is not seen in the LASCO data, it is likely a transient event similar to the pseudostreamer helical jet discussed in P. Romano et al. (2025). A faint outflow is visible at the SE limb of the Sun, even though the coronal brightness there is low. A structure resembling a polar plume near the southern solar pole shows a large transverse motion, and finally regions with radial outflows are seen in the southwestern and northwestern position angles above the Sun.

We first measure the apparent motion of the star ς Psc image using a cross-correlation routine to track its position relative to the Sun center in our image sequence. The cross-correlation analysis shows an average speed of 0.043 ± 0.003 arcseconds per second as compared with the expected ephemeris motion of 0.041 arcseconds per second. The measured direction of motion is towards an equatorial coordinate position angle of 66.8 ± 1.6 degrees, compared to the ephemeris direction of 67.8 degrees. The tight correlation between our measured and ephemeris values adds confidence to both our image alignment techniques and cross-correlation tracking methods.

Downflowing coronal features, often appearing as dark voids, have been observed in regions of solar flares (D. E. McKenzie & H. S. Hudson 1999, L. P. Li et al. 2016) moving with speeds of roughly 50 - 400 km $s^{-1}$ and showing decelerations of about 300 m $s^{-2}$. In the polar regions of the solar corona, slower downflows have also been seen in white light measurements (N. R. Sheeley & Y. -M. Wang 2002, A. N. Zhukov et al. 2026) at heights from 1.8 to 5 solar radii. Speeds of between 20 - 150 km $s^{-1}$ have been observed, and both accelerations and decelerations have been measured. In our DEB image sequence we observe a dark feature near the solar north pole at a position angle of 16 degrees. Using height vs time measurements, we track the motion of this feature from about 2.12 to 1.97 solar radii. The feature shows an average downward speed of 37.2 ± 2.9 km $s^{-1}$ and a deceleration of 13.7 ± 2.7 m $s^{-2}$. An F-test comparing the variances of the linear fit against the quadratic fit reveals F ≈ 30 which equates to less than a 1% chance that the linear fit provides as good a statistical significance as that provided by a quadratic fit. Further, this speed and deceleration agree very well with the value found by Sheeley & Wang at the lower edge of the LASCO C2 field of view (see their third and fourth figures). Sheeley & Wang propose a set of possible sources for the deceleration of these features including an outward pressure gradient, solar wind drag force, or magnetic recombination.

The radial outflows visible in the image sequence involve low contrast features and intermittent observation times and are not well-suited for using cross-correlation tracking techniques. Optical flow techniques have been used in studies of the solar corona (i.e. S. F. Gissot & J. -F. Hochedez (2007)) and here we apply the method of G. Farneback (2003) to measure the coronal flows. We compute the average flows using large box sizes of 30 x 30 image pixels, and we use all image combinations rather than just sequential images. This approach reduces noise in the average speed calculations but has lower sensitivity to accelerations. To check the result of the optical flow calculations, we first compare the optical flow speed and direction obtained for ς Psc in the coronal frames with the values measured by our cross-correlation analysis. With optical flow we measure a speed of 0.046 arcseconds per second and an average direction of 86 degrees position angle with respect to north declination. In this case the optical flow speed is 7% higher, and the direction of motion is 18.5 degrees different from the values obtained with the cross-correlation analysis. We attribute these differences to the large size of the analysis box compared to the small stellar image. For the larger coronal downflow structure in the north polar region we measure a radial inflow speed of 39.7 km $s^{-1}$ which is 6% larger than the speed measured from our height-time analysis, but within the measured error bars of the height vs time technique; 37.2 ± 2.9 km $s^{-1}$. From these checks we have confidence that the optical flow technique is returning reliable results.

Several authors have produced MHD models which use measured photospheric magnetic fields to predict the structure and dynamics of the solar corona during the 2024 April 8 total solar eclipse, including C. Downs et al. (2025), T. Baratashvili et al. (2026), and Y. Li et al. (2026). Downs et al. found good agreement with their models compared with eclipse white light images. Baratashvili et al. compared their predictions with eclipse white light and COR2 polarized brightness and found good agreement, and the models from Li et al. compared favorably with eclipse white light and Fe XIV observations. Only Li et al. published predicted maps of the coronal radial velocity (see their Figure 8a) but until now there have been no radial velocity observations to compare against their predictions. Our time sequence can fill this gap. There is a difference between the predicted velocity shown by Li et al. and our measurements; Li et al. present a slice of the predicted velocity in the meridional plane and our measurements integrate along the entire line of sight. This difference likely introduces disagreements with our measurements, but we make a first start with a qualitative comparison.

The maximum radial outflow average speed that we measure across our time sequence is 105 km $s^{-1}$ above the eastern solar limb (solar PA about 90 degrees). The position of the flow does not align with the predicted maximum radial speed from Y. Li et al. (2026), who predict a fast outflow of 200 km $s^{-1}$ above the south-western solar limb. This eastern limb flow is likely the result of a transient event, and the simulation from Li et al. is data-constrained and is not designed to capture transient events. Our radial velocity measurements do show some significant agreements with the Li et al. model. The polar downflow with an average radial velocity of $-37 \pm 3$ km $s^{-1}$ agrees in position and speed with the downflow they predict at the pseudo-streamer 1 near the north solar pole. Their predicted mixture of outflows and downflows at low heights is seen, as well as outflows in the south-western and north-western regions of the corona. Note that the color scale on their Figure 8 is different from our scale on Figure 4. In summary, we find good qualitative agreement with the Li et al. predicted radial flow field, but note that an improved flow model incorporating line of sight integration would likely show better agreement with our observations.

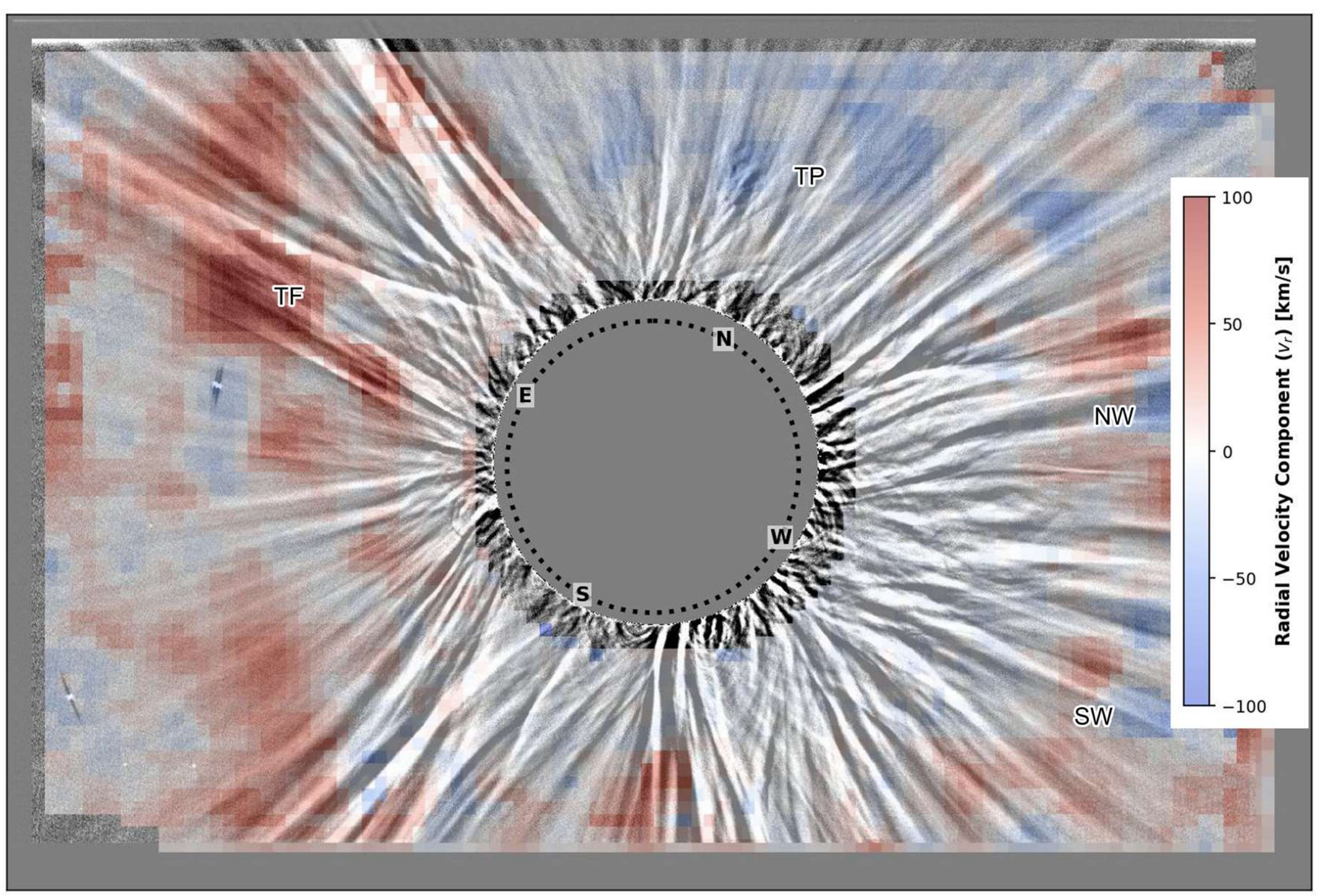

Figure 4. Radial component of plasma velocity. The radial component of plasma velocity shown here is mapped using an optical flow analysis of the DEB image sequence. The solar corona appears as a background gray scale image, and the radial velocity is not calculated in a region near the lunar limb nor near the edges of the field of view. This new type of observation can be compared with the predicted radial flow from coronal models such as Y. Li et al. (2026). A qualitative comparison with that work shows agreements with the observed north polar downflow (TP), the mixture of flows in the inner regions of this figure, and the outflows in the SW (SW) and NW (NW) regions. The strong flow above the E limb (TF) is likely from a transient flow event not addressed by the model.

## 5. Acknowledgments

The DEB Initiative acknowledges NASA award 80NSSC23K0948 and NSF award 2215167. The authors are indebted to the hard work and dedication of the more than 200 additional volunteer members of the full DEB Initiative team. Without their efforts this project would not have succeeded. The DEB Initiative team is listed here: https://debinitiative.org/main/the-deb-team/ MJP acknowledges very helpful and inspiring conversations with Jonathan Hill.